\documentclass[a4paper,twocolumn]{revtex4}

\usepackage{hyperref}
\usepackage[dvips]{graphicx}
\begin{document}

\title{Digital holography with ultimate sensitivity}

\author{M. Gross and M. Atlan }

\address{Laboratoire Kastler-Brossel de l'{}\'Ecole Normale Sup\'erieure, CNRS UMR 8552
- Universit\'e Pierre et Marie Curie, 24 rue Lhomond 75231 Paris
cedex 05. France\\
}

\begin{abstract}
We propose a variant of the heterodyne holography scheme, which
combines the properties of off-axis and phase-shifting holography.
This scheme makes it possible  filter-off numerically the zero
order image alias, and the technical noise of the reference. It is
then possible to record and reconstruct holographic images at an
extremely low signal level. We show experimentally that the
sensitivity of the method is limited only by the quantum nature of
photons.
\end{abstract}



\maketitle

Digital holography is a fast-growing research field that has drawn
increasing attention \cite{Schnars_2002}. The main advantage of
digital holography is that, contrarily to holography with
photographic plates \cite{Gabor49}, the holograms are recorded by
a CCD and the image is digitally reconstructed by a computer,
avoiding photographic processing \cite{Goodmann_1967}. Off-axis
holography \cite{Leith65} is the oldest and the simplest
configuration adapted to digital holography
\cite{Schnars_Juptner_94, Schnars94, Kreis88}.

In off-axis digital holography, as well as in photographic plates
holography, the reference or local oscillator (LO) beam is
angularly tilted with respect to the object observation axis. It
is then possible to record, with a single hologram, the two
quadratures of the object complex field. However, the object field
of view is reduced, since one must avoid the overlapping of the
image with the conjugate image alias \cite{Cuche00}.

Phase shifting digital holography \cite{Yamaguchi1997,
Leclerc2000} records several images with different phase for the
LO beam. It is then possible to obtain  the two quadratures of the
field in an on axis configuration even though  the conjugate image
alias and  the true image overlap, because  aliases can be removed
by taking  images differences.

In this paper, we propose a  digital holography technique that
combines off-axis geometry (introduced by Schnars et al.
\cite{Schnars94}), with the use of a sequence of images obtained
with different phase shifts of the LO beam to record the hologram
in amplitude and phase (as proposed by  Yamaguchi et al.
\cite{Yamaguchi1997}). To get precise phase shift,  Le Clerc's et
al. \cite{Leclerc2000} heterodyne technique is used. Using a
spatial filtering method (Cuche's et al. \cite{Cuche00}), the zero
order image, and the noise attached to it, is filtered-off
numerically. As shown experimentally, this combination of
techniques makes it possible to record and reconstruct holographic
images at a very low level of signal: $1$ photo electron of signal
per reconstructed image pixel during a whole sequence of 12 images
($\simeq 1$ s). This corresponds to the ultimate quantum limit.


\begin{figure}[]
\begin{center}
\includegraphics[width = \linewidth,keepaspectratio=true]{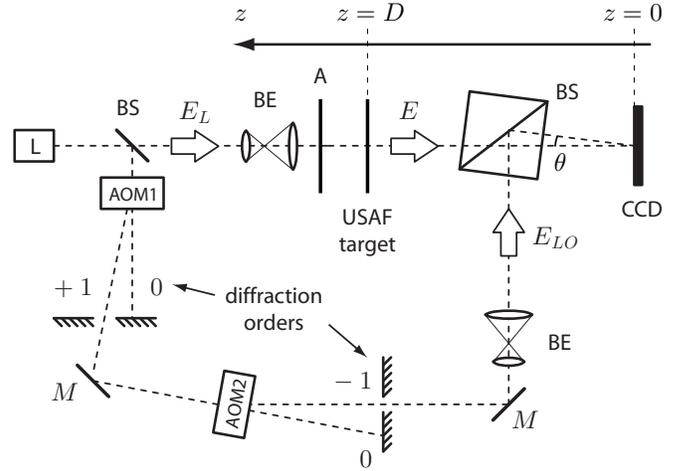}
\caption{ USAF target digital holography setup. L: main laser; BS:
Beam splitter; AOM1 and AOM2: acousto optic modulators; BE: beam
expander;  M: mirror;  A: light attenuator. USAF: transmission
USAF target. CCD : CCD camera. } \label{fig_setup_usaf}
\end{center}
\end{figure}

The setup is shown on Fig.\ref{fig_setup_usaf}. It is similar to
the one used in \cite{Leclerc2000,Leclerc2001}. The main laser L
is a Sanyo DL-7140-201 diode laser ($\lambda=785 \, \rm  nm$, $50
\, \rm  mW$ for $95 \, \rm mA$ of current). It is split into an
illumination beam (frequency $\omega_{L}$, complex field $E_{L}$),
and in a LO beam ($\omega_{LO}$, $E_{LO}$). The object we want to
image is an USAF target in transmission, which is back
illuminated. The object signal field $E$ is not  shifted in
frequency ($\omega_L$). A set of  optical attenuators A (grey
neutral filter) is used  to reduce the illumination. The CCD
camera (PCO Pixelfly digital camera: $12  \, \rm  bit$, frame rate
$\omega_{CCD}=12.5  \, \rm Hz$, acquisition time $T=1/12.5=80 \,
\rm  ms$, with $1280 \times 1024$ pixels of $6.7 \times 6.7  \,
\rm \mu m$) records the hologram of the object, i.e. the object
($E$) and LO ($E_{LO}$) field interference pattern. Using AOM1 and
AOM2 acousto optic modulators (Crystal Technology:
$\omega_{AOM1,2} \simeq 80  \, \rm MHz$), the optical frequency
$\omega_{LO}$ of the LO beam, can be freely adjusted. To make
4-phases detection of the object field $E$, the LO frequency is
adjusted to be $\omega_{LO}=\omega_L + \omega_{CCD}/4$
\cite{Leclerc2000}. Moreover the LO beam is tilted (angle $\theta
\sim 1 ^\circ$) with respect to the camera to object observation
axis. Our holographic setup thus works both in off-axis and in
phase shifting mode.

A sequence of 12 CCD images $I_0$ to $I_{11}$ (measurement time
$0.96 \, \rm s$) is recorded. Since the LO beam is phase shifted
by $\pi/2$ between two consecutive images, the object complex
hologram $H$ is obtained by summing the CCD images with the
appropriate phase shift:
\begin{equation}\label{equ_holo}
    H= \sum_{m=0} ^{11}  (j)^m I_m
\end{equation}
where $m$ is the image index and $j^2=-1$.

We have reconstruct the images of the USAF target by using the
standard convolution method \cite{Schnars94,Kreis2000} that yields
a calculation grid equal to the pixel size. To calculate the
convolution product, we have used the Fourier method, like in
\cite{Leclerc2000}. To make faster Fourier calculation the $1028
\times 1024$ CCD data are truncated to a $1024 \times 1024$ 2D
matrix.

The reconstruction algorithm is the following. The real space hologram
$H(x,y,z)$ in the $z=0$ CCD plane is calculated by Eq.\ref{equ_holo}.
The hologram in the CCD reciprocal plane (i.e. in the $z=0$ k-space) is
obtained by Fourier transformation:
\begin{equation}\label{Eq_image_H(0)}
  \tilde H(k_x,k_y,0)=FFT \left[H(x,y,0)\right]
\end{equation}
The k-space hologram at any distance $z$ from the CCD is then:
\begin{equation}\label{Eq_image_H(x)}
  \tilde H(k_x,k_y,z)= \tilde K(k_x,k_y,z) \tilde H(k_x,k_y,z)
\end{equation}
where $\tilde K$ is the k-space kernel function that describe the
propagation from $0$ to $z$.
\begin{equation}\label{equ_K_kernel}
    \tilde K(k_x,k_y,z)=e^{j z ({k_x}^2 + {k_y}^2 )/k}
\end{equation}
where $k=2\pi/\lambda$ is the optical wave vector. The
reconstructed image, which is the hologram in the object plane
($z=D$), is then obtained by reverse Fourier transformation:
\begin{equation}\label{Eq_image_H(z)}
  H(x,y,z)=FFT^{-1} \left[\tilde H(k_x,k_y,z)\right]
\end{equation}
\begin{figure}[]
\begin{center}
\includegraphics[width = \linewidth,keepaspectratio=true]{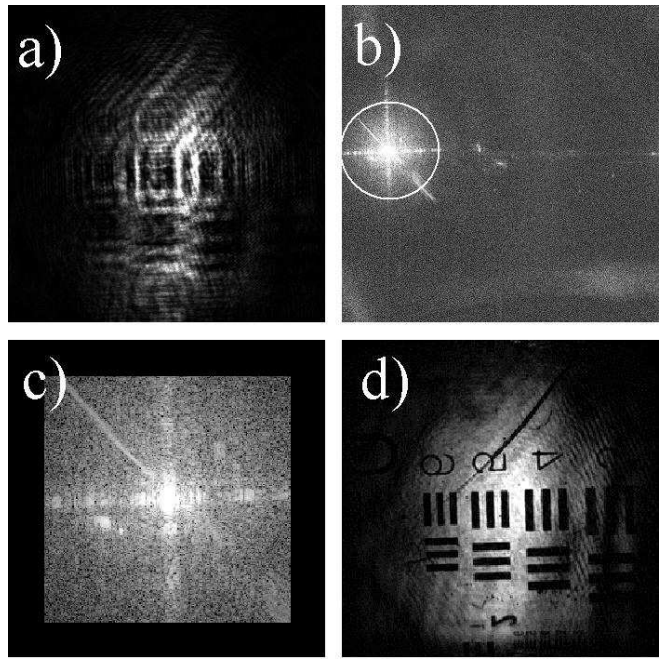}
\caption{Reconstruction of the USAF target image. (a) Intensity
image of the  CCD complex hologram $|H(x,y,0)|^2$ in linear grey
scale. (b) Intensity image of the k-space hologram $|\tilde
H(k_x,k_y,0)|^2$ in logarithmic grey scale. (c) Intensity image of
the truncated k-space hologram in logarithmic grey scale. (d) USAF
target reconstructed image, i.e.  intensity image of the real
space hologram in the object plane $|H(x,y,D)|^2$ in linear grey
scale. Images are $1024 \times 1024$ (a,b), or in $256 \times 256$
(c,d) pixels.} \label{fig_ima_4ph_700}
\end{center}
\end{figure}
Fig.\ref{fig_ima_4ph_700}a shows the intensity of the CCD plane
complex hologram (i.e. $|H(x,y,0)|^2$). The USAF target is seen,
but is blurred.  Fig.\ref{fig_ima_4ph_700}b shows the intensity of
the CCD plane k-space hologram (i.e. $|H(k_x,k_y,0)|^2$). The
bright zone in the left hand side of the image corresponds to the
true holographic image i.e. to the interference of the object
field with the LO field. The zero-order and twin image
\cite{Cuche00} alias contributions are very low. The zero-order
alias, in the center of the k-space hologram (pixel $512,512$), is
weak, because the contribution of the LO intensity cancels out by
making difference of images (Eq.\ref{equ_holo}). Since the phase
shift provided by the acousto optic modulators is very close to a
multiple of $\pi/2$, the twin-image is also very weak. It is
barely not visible on Fig.\ref{fig_ima_4ph_700}b, although the
display is in grey logarithmic scale.

To select the relevant first order image information, and to fully
suppress the zero-order and twin image aliases, we have used the
k-space filtering method developed by Cuche et al. \cite{Cuche00}.
We have selected, in the k-space $1024\times 1024$ matrix, a $200
\times 200$ region of interest centered on the true image bright
zone. Note that this selection is made possible by the off-axis
geometry that has translated the true image in the left hand side
of k-space matrix. The selected area is then copied in the center
of a $256 \times 256$ zero matrix (zero padding). The calculation
of the $z=D$ k-space and real space holograms
(Eq.\ref{Eq_image_H(x)} and Eq.\ref{Eq_image_H(z)}) are thus done
on this $256 \times 256$ calculation grid.
Fig.\ref{fig_ima_4ph_700}c shows the object plane k-space hologram
(i.e. $\tilde H(k_x,k_y,z)$) in the $256 \times 256$ grid.
Fig.\ref{fig_ima_4ph_700}d shows the object plane real space
hologram (i.e. $H(x,y,z)$) obtained by the Eq.\ref{Eq_image_H(z)}
reverse FFT, which is the reconstructed image of the USAF target.
Note that the translation of the selected zone in the center of
the k-space in Fig.\ref{fig_ima_4ph_700}c moves the reconstructed
USAF target to the center of the image as seen in
Fig.\ref{fig_ima_4ph_700}d.

We have assessed the sensitivity limit of this off-axis heterodyne
holography technique by recording images of  the USAF target at
different levels of illumination. To get quantitative results, we
have determined  the absolute number of photo electrons that
correspond to the signal beam. First, we have calibrated our set
of attenuators (A) at the working wavelength (785 nm). These
attenuators were then used to change the illumination of the USAF
target while the laser power remained constant throughout the
experiment. For a high level of illumination of the USAF target,
and without LO beam, we have measured the signal beam in photo
electrons units: we have multiplied the CCD signal in digital
counts (0...4095 DC) by  the CCD gain ($G=2.4$ photo electrons per
DC: PCO calibration). For other levels of illumination (for a
given set of attenuators in the object arm), we have calculated
the signal beam level in photo electrons units, from the
attenuators calibration.

\begin{figure}[]
\begin{center}
\includegraphics[width = \linewidth,keepaspectratio=true]{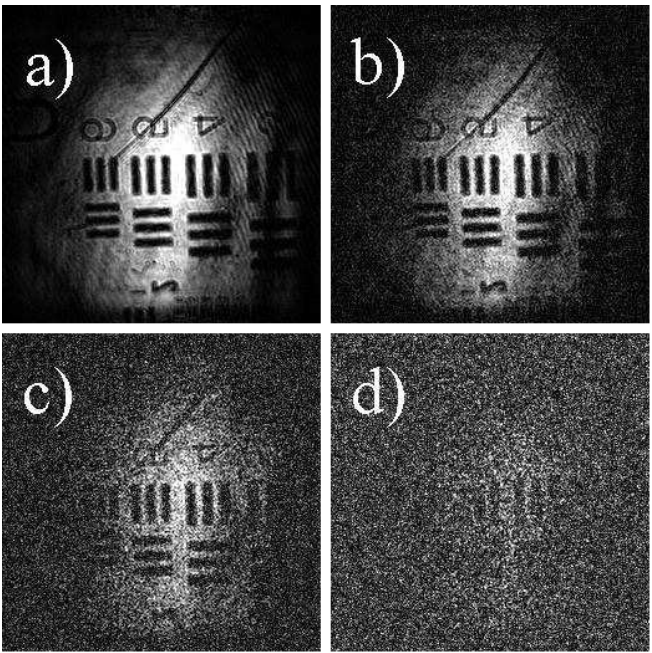}
\caption{Reconstructed image of a USAF target in transmission with
low light illumination. Images are obtained with a signal field
that corresponds to $\simeq 2.8 \times 10^7$ (a), $2.8\times 10^5$
(b), $4\times 10^4$ (c) and $6000$ (d) photo electrons for the
whole 12 CCD images ($256 \times 256$ pixels displayed in linear
grey scale). } \label{fig_usaf_4ph}
\end{center}
\end{figure}

Fig.\ref{fig_usaf_4ph} shows the reconstructed images that are
obtained at various levels of illumination.
Figs.\ref{fig_usaf_4ph} a to d are obtained with $\simeq 2.8
\times 10^7$ (a), $2.8\times 10^5$ (b), $4\times 10^4$ (c)  and
$6000$ (d) photo electrons of signal for the whole set of 12 CCD
images. It should be noticed that the total number of photo
electron is not the relevant parameter to characterize the image
signal. Since the reconstruction is performed with a truncation
over a $200 \times 200$ pixels region of the k-space, the
reconstructed image has about $4\times 10^4$ resolved pixels. The
signal can thus be characterized by the number of photo electrons
per resolved pixel. Therefore, the images of
Fig.\ref{fig_usaf_4ph} correspond to $700$ (a), $7$ (b), $1$ (c)
and $0.15$ (d) photo electrons per pixel. For comparison, the
image of Fig.\ref{fig_ima_4ph_700}d corresponds to $3.5 \times
10^4$ photo electrons per pixel.

One photo electron per resolved pixel is the quantum limit of
visibility of an image (i.e. the limit that is related to the
quantum character of photons, and photo electrons).
Fig.\ref{fig_usaf_4ph}c image corresponds exactly to this limit,
where the SNR of the image is about 1. This demonstrate that our
off-axis heterodyne technique is able to filter-off technical
noise and to reach the quantum limit of photodetection. Note that
it is still possible to perceive the USAF target on
Fig.\ref{fig_usaf_4ph}d with only $0.15$ photo electron per pixel
(SNR$\simeq 0.15$).

This result can be understood quite simply. In the case of a very
weak signal, the majority of the technical noise is within the LO
beam (because the signal beam is much weaker than the LO beam).
This noise can be eliminated  by a double filtering process.
First,  most of the noise is removed by taking the difference of
successive images (Eq.\ref{equ_holo}). This is the phase-shifting
filtering that occurs in the temporal frequency domain. But this
method doesn't remove all the technical noise. Since the LO beam
extends within a few spatial modes (i.e. a few pixels of the
k-space), its noise is located near the center of the k-space.
This means that the LO technical noise is flat field (it varies
slowly in space). A spatial filtering similar to the one
introduced by Cuche et al. \cite{Cuche00} removes the rest of the
technical noise which lies in the center of the k-space. This is
the off-axis filtering that occurs in spatial frequency domain.

The digital holography technique presented here relies on the
combination of off-axis and phase-shifting configurations to
record holograms with ultimate sensitivity. This technique lets
one  filter-off the local oscillator technical noise and  reach
the quantum sensitivity limit. As demonstrated, we were able to
make an image with a signal of one photoelectron per pixel during
a sequence of 12 recorded camera frames. Although demonstrated in
through transmission this technique is expected to work as well in
reflection with diffusely scattering surface. It might be used to
perform holography with at extremely low signal level, e.g.
nano-object imaging... We have used a variant of this technique to
detect the so called "tagged photons" in acousto-optic imaging
\cite{Gross_03}, or the photons that are transmitted through the
breast, in vivo \cite{Gross_05}. With a high quantum efficiency
camera, the technique could also be used to perform quantum optics
tests (non classical photon statistics, squeezed states).

The authors acknowledge the French ANR for its support.

\pagebreak

\end{document}